\documentclass[useAMS,usenatbib]{mnras}
\def\draftversion{0} 
\usepackage{amssymb,latexsym,graphicx,natbib,eufrak,times,amsmath,xspace,ifthen}

\ifthenelse{ \draftversion > 0 }{
  \newcommand{\new}[1]{\textbf{#1}} 
} {
  \newcommand{\new}[1]{#1}
}

\ifthenelse{\equal{\draftversion}{1}}{
  \usepackage{xcolor}
  \newcommand{\sep}[1]{\par\begin{color}[rgb]{0,0.4,0}\begin{center}\hrule\end{center}\end{color}\par} 
  \newcommand{\todo}[1]{\begin{color}{red}\ \ifthenelse{\equal{#1}{}} {$\bullet\bullet\bullet$} {$\bullet$\ #1 $\bullet$}\end{color}} 
  \newcommand{\idea}[1]{\begin{color}[rgb]{0,0.4,0}\textit{#1}\end{color}} 
  \newcommand{\sk}[1]{\begin{color}[rgb]{0.6,0,0.6}#1\end{color}} 
  \newcommand{\toc}{\par\begin{color}[rgb]{0.6,0,0.6}\begin{center}\hrule\vspace{0.5mm}\begingroup\small\let\cleardoublepage\relax\let\clearpage\relax\mytoc\endgroup\vspace{0.5mm}\hrule\end{center}\end{color}\par} 

  }{
  \newsavebox{\trashcan}

  \newcommand{\sep}[1]{}
  \newcommand{\todo}[1]{}
  \newcommand{\idea}[1]{}
  \newcommand{\sk}[1]{}
  \newcommand{\toc}{}

  }
 \makeatletter\newcommand\mytoc{\@starttoc{toc}}\makeatother 
\long\def\symbolfootnote[#1]#2{\begingroup%
\def\thefootnote{\fnsymbol{footnote}}\footnote[#1]{#2}\endgroup} 

\newcommand{\fig}[2][]{Figure#1~\ref{fig:#2}}

\newcommand{\sect}[2][]{Section#1~\ref{sec:#2}}

\newcommand{\bb}[1]{\ifmmode \mbox{\boldmath $ #1$} \else  \mbox{\boldmath $#1$} \fi}

\newcommand{\mh}{\ensuremath{\textrm{\,--\,}}}    
\newcommand{\U}[1]{\ensuremath{\mathrm{~#1}}}     
\newcommand{\e}[1]{\ensuremath{\times 10^{#1}}}   

\newcommand{\yr}{\U{yr}}
\newcommand{\Myr}{\U{Myr}}          
\newcommand{\Gyr}{\U{Gyr}}          
\newcommand{\pc}{\U{pc}}
\newcommand{\kpc}{\U{kpc}}
          
\newcommand{\Msun}{\U{M}_{\odot}}   
\newcommand{\Msunyr}{\Msun\yr^{-1}} 
   
\newcommand{\cc}{\U{cm^{-3}}}

\newcommand{\dex}{\U{dex}}

\newcommand{\afe}{\ensuremath{[\alpha/\mathrm{Fe}]}\xspace}       
\newcommand{\feh}{\ensuremath{[\mathrm{Fe/H}]}\xspace}       

\newcommand{\ramses}{{\small RAMSES}\xspace}

\newcommand{\vintergatan}{{\small VINTERGATAN}\xspace}

\newcommand{\lund}{Department of Astronomy and Theoretical Physics, Lund Observatory, Box 43, SE-221 00 Lund, Sweden}
\newcommand{\surrey}{Department of Physics, University of Surrey, Guildford, GU2 7XH, UK}

\defcitealias{Agertz2020}{Paper I}
\defcitealias{Renaud2020}{Paper II}

\title[VINTERGATAN III]{VINTERGATAN III: how to reset the metallicity of the Milky Way}
\author[Renaud et al.] {Florent~Renaud$^1$\thanks{florent@astro.lu.se}, Oscar~Agertz$^{1}$, Eric~P.~Andersson$^{1}$, Justin~I.~Read$^{2}$, Nils~Ryde$^1$, \newauthor
Thomas~Bensby$^{1}$, Martin~P.~ Rey$^{1}$ and Diane~K.~Feuillet$^{1}$\\
$^1$ \lund\\
$^2$ \surrey}

\date{Accepted 2021 February 22. Received 2021 January 18; in original form 2020 June 12}

\begin{document}
\maketitle


\begin{abstract}
Using the cosmological zoom simulation \vintergatan, we present a new scenario for the onset of star formation at the metal-poor end of the low-\afe sequence in a Milky Way-like galaxy. In this scenario, the galaxy is fueled by two distinct gas flows. One is enriched by outflows from massive galaxies, but not the other. While the former feeds the inner galactic region, the latter fuels an outer gas disk, inclined with respect to the main galactic plane, and with a significantly poorer chemical content. The first passage of the last major merger galaxy triggers tidal compression in the outer disk, which increases the gas density and eventually leads to star formation, at a metallicity 0.75 dex lower than the inner galaxy. This forms the first stars of the low-\afe sequence. These in situ stars have halo-like kinematics, similarly to what is observed in the Milky Way, due to the inclination of the outer disk which eventually aligns with the inner one via gravitational torques. We show that this tilting disk scenario is likely to be common in Milky-Way like galaxies. This process implies that the low-\afe sequence is populated in situ, simultaneously from two formation channels, in the inner and the outer galaxy, with distinct metallicities. This contrasts with purely sequential scenarios for the assembly of the Milky Way disk and could be tested observationally.
\end{abstract}

\section{Introduction}

Understanding the formation and evolution of galaxies, in particular of the Milky Way, is one of the biggest challenges of modern astrophysics. A natural approach to this task is to connect events and processes along galaxy evolution to the observable signatures they leave on the properties of the stellar populations at the present-day. For the Milky Way, this is made possible by the accumulation of multi-dimensional observational data by surveys like APOGEE \citep{Hayden2014, Anders2017}, LAMOST \citep{Xiang2017}, GALAH \citep{Lin2019, Buder2019} and Gaia \citep{Gaia2018}. Combining multiple indicators like chemistry, ages and up to 6D phase-space allows us to propose or rule out formation scenarios for stellar sub-populations as well as the entire Galaxy.

For instance, in the \feh-\afe plane, stars gather in two sequences, one at high and one at low \afe (e.g. \citealt{Fuhrmann1998, Anders2014, Nidever2014, Hayden2015}). These two groups are approximately associated with the geometrical distinction of the respectively thick and thin disks (\citealt{Gilmore1983, Prochaska2000, Bovy2016}), kinematically hot and cold \citep{Hayden2020}, and old ($\gtrsim 10 \Gyr$) and young populations \citep{Fuhrmann2011, Schuster2012, Bensby2014, Feuillet2019}. Yet, the origins of these two populations and the transition from one to the other remain to be fully understood, and the factual correspondence between the chemical, geometrical, kinematic and age properties needs to be established quantitatively and theoretically \citep[see e.g.][]{Kawata2016}.

To explain the main observed features, the two-infall scenario (\citealt{Chiappini1997}, revised in \citealt{Spitoni2019}), proposes that the galactic stellar population is assembled in a self-enriching interstellar medium (ISM, i.e. with increasing metallicity) along the high-\afe sequence, until the accretion of low metallicity gas dilutes the ISM of the galaxy. The global metallicity of the gas is then reduced to form the metal-poor end of the low-\afe sequence which subsequently enriches to form the entire sequence \new{(see a similar approach in \citealt{Lian2020})}. In this \emph{sequential} model, it is thus the second infall of gas which triggers the transition from the high- to the low-\afe branch.

By contrast, in the scenario from \citet{Clarke2019}, the decrease of the surface density of star formation rate (SFR) in massive gas clumps is invoked to explain the bimodality in \afe, without requiring any external factors. It is likely that such an intrinsic evolution of the star formation conditions is a consequence of the decrease of the gas fraction in disks, and a change in the instability regime \citep[][Renaud et al. in preparation]{Agertz2015b, Romeo2020}. Therefore, the model by \citet{Clarke2019} also depicts the chemical assembly of the galaxy as a sequential process. In their disks simulated in isolation, \citet{Khoperskov2020} found that the chemical bimodality arises from an intrinsic evolution of the star formation activity, and that the metallicity spread of the low-\afe sequence originates from radial gradients. While such models successfully reproduces the overall observed distributions in chemical space \citep[see also][]{Amarante2020, Beraldo2020}, the impact of the missing cosmological context on the results remains uncertain. 

By combining APOGEE DR14 \citep{Abolfathi2018} and Gaia DR2 data \citep{Gaia2018}, \citet{Feuillet2019} established that, in the solar neighborhood, the highest metallicity stars (\feh $\gtrsim 0.2 \dex$) are older than stars at solar metallicity. They suggest that the metal-rich end of the low-\afe sequence is the continuation of the high-\afe sequence in the inner disk, before the onset of the low-\afe sequence itself. Efficient radial migration of chemically enriched stars from the inner galaxy to the solar neighborhood would then explain that solar metallicity stars are younger than stars at higher metallicities (see also \citealt{Miglio2020}). \citet{Ciuca2020} confirmed their findings by improving relative age estimates using the machine learning technique on APOGEE data. In addition, they highlighted that stars in the outer disk lie at the low-metallicity end of the low-\afe sequence, while stars in the inner disk follow the high-\afe sequence (see also {\citealt{Bensby2010, Bensby2011}).

Both \citet{Feuillet2019} and \citet{Ciuca2020} reported that the metal-rich and metal-poor ends of the low-\afe sequence are coeval. Therefore, a single, continuous evolutionary track in the \feh-\afe plane cannot represent the entirety of the stellar population. This rules out sequential formation scenarios and calls for multiple star formation channels to be simultaneously active $\approx 9.6 \mh 9.8 \Gyr$ ago to explain the detection of chemically distinct populations of the same age.

A natural explanation consists in invoking the superposition of two independent evolutionary tracks, from two distinct galaxies that eventually merge. Stars at the metal-poor extremity of the low-\afe sequence would then be of accreted origin, likely from a low-mass satellite which did not self-enrich as efficiently as the Milky Way. Indeed, low-\afe stars at $\feh \lesssim -1 \dex$ (i.e. at a much lower metallicity than the bulk of the thin disk) show halo-like kinematics, with a significant fraction found on retrograde orbits \citep{Haywood2018}. \citet{diMatteo2019} argues that these stars were accreted during a galaxy merger that also dynamically heated the pre-existing galactic disk $\approx 9\mh 11 \Gyr$ ago. In that case, the accreted material would account for the dynamically hot component and retrograde motions. The heated thick disk would then constitute the only in situ component of the kinematically-defined halo. However, it is yet unclear whether, to what extent, and how, this very low-metallicity component connects to the low-\afe sequence of the Milky Way at higher metallicities ($\feh \gtrsim -0.5 \dex$).

In this paper, we present the tilting disk scenario, an explanation for the formation of the metal-poor end of the low-\afe sequence. This process plays a central role in the onset of the \afe bimodality, and the formation of the radially-extended thin disk. We use the \vintergatan simulation (a cosmological zoom simulation of a Milky Way-like galaxy introduced in \citealt{Agertz2020} and \citealt{Renaud2020}, hereafter \citetalias{Agertz2020} and \citetalias{Renaud2020} respectively)\footnote{Movies are available at:\\ \url{http://www.astro.lu.se/~florent/vintergatan.php}} to trace these stars back to their birth epoch and sites, and describe their formation mechanism. The details of the simulation are briefly summarized in \sect{method}. \sect{results} presents the processes leading to the formation of these stars. We discuss the predictions and the likelihood of this scenario in \sect{discussion}, and summarize our findings in \sect{conclusion}.

\section{Method}
\label{sec:method}

We use the \vintergatan cosmological zoom simulation of a Milky Way-like galaxy. Numerical details and the overall description of the method are presented in \citetalias{Agertz2020} and only briefly summarized below. 

The simulation is run with the adaptive mesh refinement code \ramses \citep{Teyssier2002}. The initial conditions are that of the ``m12i'' halo from \citet{Hopkins2014}, re-simulated from $z=100$ at a maximum resolution of $\approx 20 \pc$. The model of \citet{Haardt1996} describes heating from the UV background. Cooling is metallicity-dependent, using the tabulated functions of \citet{Sutherland1993} and \citet{Rosen1995}. Star formation proceeds in the gas denser than $100 \cc$ with the local efficiency prescription of \citet{Padoan2011}. Stellar feedback in the form of winds, radiation pressure, type-II and type-Ia supernovae (SNe) is included following the implementation of \citet{Agertz2015}, and accounting for the injection of mass, energy, momentum and heavy elements. The injection of momentum by SNe depends on the resolution of the local cooling radius, following \citet{Kim2015}.

Oxygen and iron injected by SNe are traced by passive scalars. The \afe and \feh abundance ratios are computed assuming solar composition \citep{Anders1989}. In our analysis, the abundance in $\alpha$ elements is solely traced by oxygen. As discussed in \citetalias{Agertz2020}, \emph{the abundances presented are raw values} from the simulation. Yields and SN rates are highly uncertain in models, which leads to mismatches between simulations and observations. We chose to not re-normalize these values at the post-processing stage and only provide the raw data. When comparing with observations or other simulations, we only refer to relative comparisons in the chemical features in \afe and \feh and not to their absolute values.

At the end of the simulation, the \vintergatan galaxy resembles the Milky Way, including a bimodality in the \afe-\feh plane. A detailed comparison with the real Galaxy is shown in \citetalias{Agertz2020} and the full formation scenario is presented in \citetalias{Renaud2020}.

\section{Results}
\label{sec:results}

\vintergatan experiences the final coalescence of its last major merger at $z = 1.2$ ($\approx 8.7 \Gyr$ ago), with a galaxy 3 times less massive. This event takes place approximately at the same epoch as the onset of the low-\afe sequence, which suggests a casual link between the two. However, because the companion galaxy has a comparable enrichment history as \vintergatan (due to their similar masses), it does not contain stars with significantly different chemical compositions (\citetalias{Renaud2020}). Therefore, the companion galaxy cannot directly contribute to a new, chemically distinct stellar population. Yet, the passage of the galaxy itself is key in altering the hydrodynamics of the ISM, and triggering the assembly of the low-\afe sequence in a structure that will become the backbone of the outer galactic disk, as described in this Section.

\subsection{Before the interaction}
\label{sec:before}

\begin{figure*}
\centering
\includegraphics[scale=0.9]{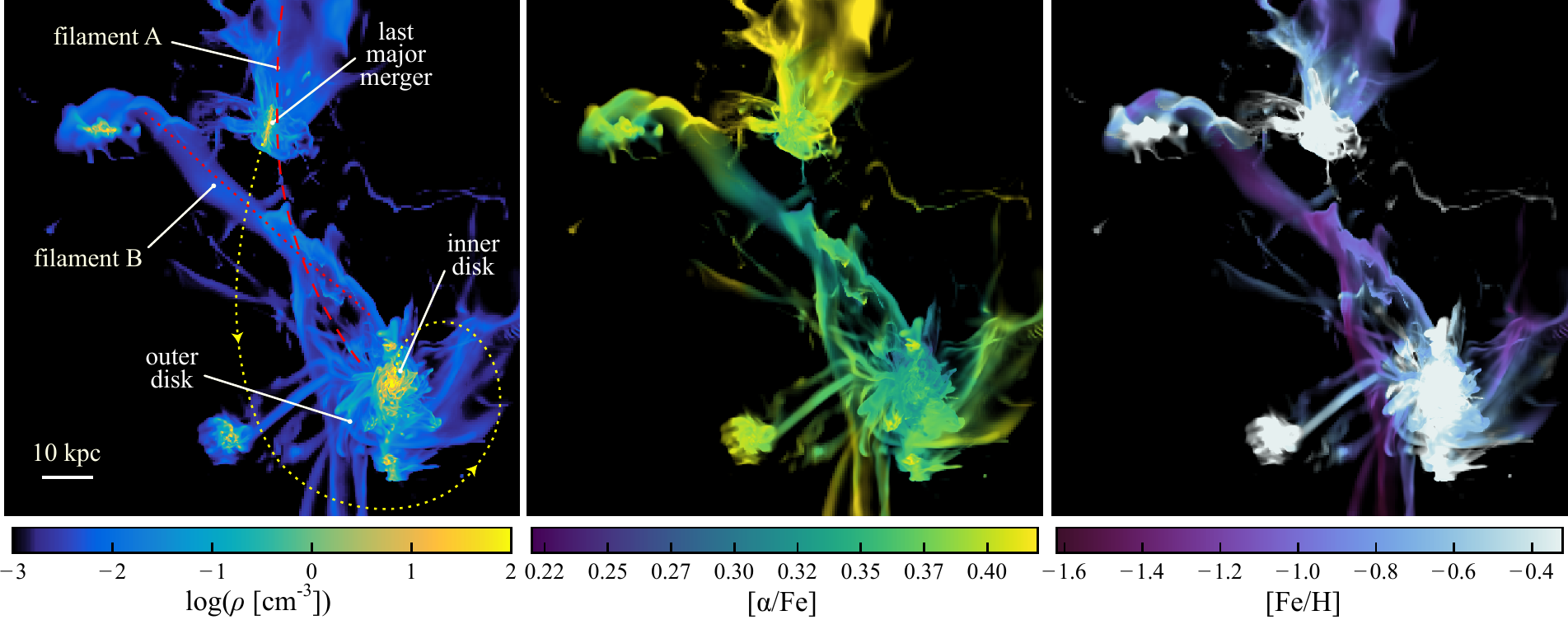}
\caption{Maps of gas density (left), \afe (center) and \feh (right) at $z\sim 1.6$, before the first passage of the last major merger. The maps are masked to highlight the dense gas only. The yellow dotted line indicates the approximate future orbit of the last major merger. The two main galaxies (\vintergatan and its last major merger) are connected by the filament A (dashed red line), building and fueling their disks. The two galaxies have progressively accreted most of the filament's content, such that the gaseous flow is discontinued in between the galaxies before the companion arrives. Conversely, the filament B (dotted red line) still fuels \vintergatan, but not in the disk plane. The filament B hosts only a low-mass galaxy (visible in the top-left corner), which had not undergone important interactions nor starbursts. Thus, the filament B has lower \afe and \feh than the filament A which has been enriched by galactic outflows from the two main galaxies.}
\label{fig:tilt}
\end{figure*}

At $z=1.6$ ($\approx 10 \Gyr$ ago), before its first encounter with the last major merger, \vintergatan is a disk galaxy of stellar mass $\approx 3\e{10} \Msun$ hosting on-going star formation up to a radius of $\approx 5 \kpc$. Therefore, the outer galactic disk and the equivalent of the solar neighborhood have not formed yet. The stars formed at this epoch have\footnote{Due to uncertainties on the yields and SN-Ia rates, the \afe in the simulation is about $0.2 \mh 0.3 \dex$ higher than observed in the solar neighborhood. See \citetalias{Agertz2020} for details.} $\afe \approx 0.35$ to $0.42 \dex$ and \feh $\approx 0$ to $0.5 \dex$. In situ formation proceeds within this rather wide metallicity range, but with a decreasing \afe along the high-\afe sequence (see \citetalias{Renaud2020}, Figure 4). Soon, this will become the junction between the high- and low-\afe sequences, corresponding to the ``bridge'' identified in observational data by \citet{Ciuca2020}. At this time, all the low-metallicity stars ($\feh \lesssim -0.5 \dex$) are found in the high-\afe sequence, having formed either $\approx 1 \Gyr$ earlier in low-mass satellites, or $\approx 2\mh 3 \Gyr$ earlier in the case of in situ stars (i.e. $\approx 11$ and $13 \Gyr$ ago, respectively, see \citetalias{Renaud2020}).

Until this epoch, the ISM of \vintergatan is mainly fueled by an intergalactic filament, labeled ``A'' in \fig{tilt}. This filament and nearby structures in the cosmic web set the angular momentum of the galactic disk and thus its orientation. Here, the cold flow reaching the galaxy is coplanar with the disk (but see \citealt{Dubois2014} for counter-examples). This filament also hosts another massive galaxy (stellar mass of $\approx 10^{10} \Msun$, i.e. $1/3$ of the Milky Way) moving towards \vintergatan, and which will become its last major merger. While fueling these two massive galaxies, the filament slowly becomes depleted in gas. The remaining medium is polluted by the outflows from both galaxies. This material mixes with the intergalactic medium of very low-metal content and, in the vicinity of \vintergatan, the resulting gas yields intermediate abundances of $\afe \approx 0.38$ to $0.45 \dex$ and $\feh \approx -1.0$ to $-0.6 \dex$ (central and right panels of \fig{tilt}). 

At the same time, \vintergatan is also fueled by a secondary filament, labeled "B" in \fig{tilt}, which comprises another but significantly less massive galaxy (stellar mass of $\approx 7\e{8}\Msun$, i.e. approximately 45 and 15 times less than \vintergatan and the last major merger, respectively). Because of its small mass, this galaxy harbors a modest star formation activity ($\approx 0.3 \Msunyr$, compared to more than $20 \Msunyr$ in \vintergatan at this epoch) and thus only yields a weak chemical enrichment of its outflowing material. Therefore, the absence of another massive polluter in the flow B implies a poorer chemical composition than in the flow A. Furthermore, the chemical enrichment of this filament occurs in a slow and smooth manner, as opposed to the bursty regime of mergers polluting the filament A. Therefore, the \afe content is also lower in the filament B than in A (\citetalias{Renaud2020}). In the vicinity of \vintergatan, we find that the chemical composition of the gas in the filament B is $\afe \approx 0.30$ to $0.32 \dex$ and $\feh \approx -1.4$ to $-1.0 \dex$, i.e. respectively $\approx 0.15 \dex$ and $\approx 0.4 \dex$ less than in the filament A, on average.

As distinct structures, filaments A and B are not coplanar. When the material from the filament B reaches \vintergatan, it forms a flattened structure in the outskirts of the galaxy, misaligned with the inner disk. Due to its angular momentum, this outer structure does not reach the inner galaxy, but rather spans radii between $5 \kpc$ and $10 \mh 15 \kpc$ (depending on the azimuth), with densities in the range of that of the circumgalactic medium ($\lesssim 10^{-2} \cc$). This tilted outer gas disk is therefore too diffuse to form stars.

\subsection{The interaction and the trigger of star formation in the outer disk}
\label{sec:during}

\begin{figure}
\centering
\includegraphics{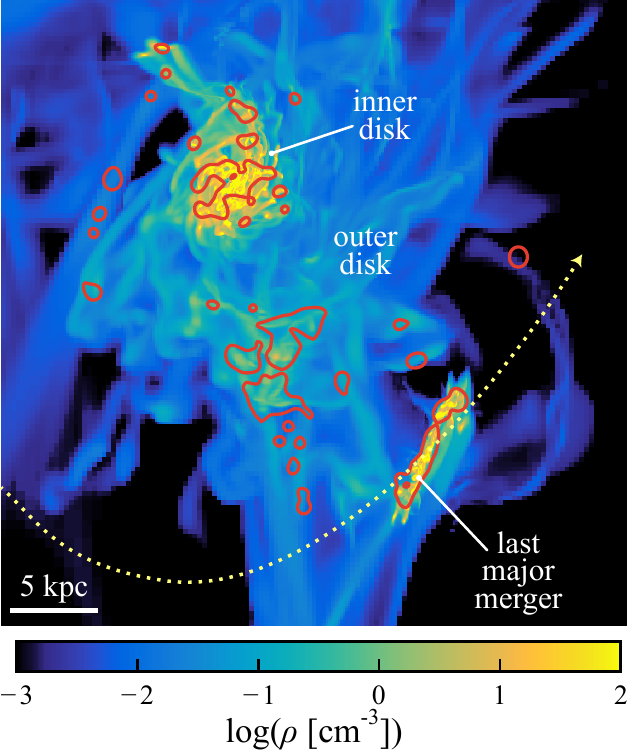}
\caption{Gas density map showing \vintergatan (top-left) and the last major merger (bottom-right, almost edge-on) during the separation of the two galaxies, about $450 \Myr$ before their merge. The last major merger is now moving toward the top of the figure. The red contours show the areas of tidal compression, occurring in several regions within the central parts of the galaxies, but also over large volumes in the outer disk \emph{between} the galaxies (spanning $\approx 10 \kpc \times 5 \kpc$ in the projection shown). The gas density is increased by several orders of magnitude in these volumes, favoring the collapse of gas clumps leading to future star formation.}
\label{fig:compress}
\end{figure}

The first encounter between \vintergatan and the last major merger occurs at $z=1.5$ ($9.6 \Gyr$ ago). The large pericenter distance ($\approx 20 \kpc$) implies that the intruder does not reach the densest parts of the main galaxy, and thus only experiences weak dynamical friction from the diffuse corona and the dark matter halo. Therefore, the intruder retains a large orbital energy, and can separate from \vintergatan for a long period before falling back. (The final coalescence will take place $\approx 900 \Myr$ later, at $z=1.2$, corresponding to $8.7 \Gyr$ ago.) This first brushing passage does not induce a strong morphological transformation of the inner \vintergatan, but significantly alters the structure of its outer titled disk, during the long interaction period.

The misalignment of the three disks (inner and outer of \vintergatan and that of the intruder) and the relatively large separation between the galaxies during the first passage do not lead to any strong spin-orbit coupling, and tidal tails are barely formed from any of these structures \citep{Duc2013}. However, the combined gravitational potential of the two galaxies modifies the nature of the tidal field and induces tidal compression on kpc-scale volumes (as in all interactions, \citealt{Renaud2009}) within the galaxies and in the outer regions, as shown with contours in \fig{compress}. By inverting the destructive effect of classical tides, tidal compression effectively increases the gas density in these regions. In \fig{compress}, a particularly large volume in between the two galaxies, $8 \kpc$ away from the center of \vintergatan, undergoes a strong compression. As the interaction proceeds, the companion galaxies faces other azimuths, such that a very large fraction of the outer disk undergoes an episode of compression before the coalescence phase. 

\begin{figure}
\centering
\includegraphics{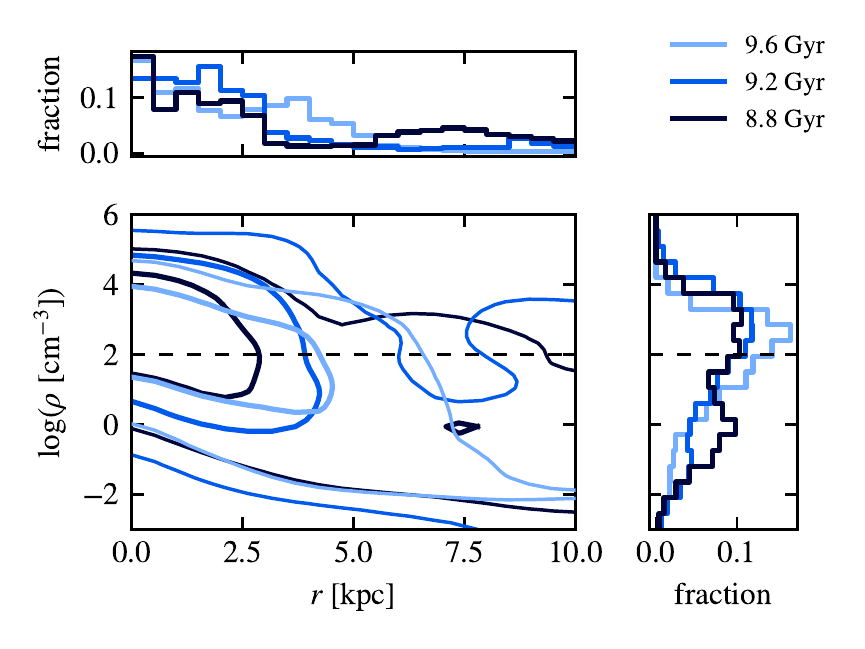}
\caption{Distribution of gas mass as a function of galacto-centric radius and gas density, before (light blue), during (medium blue) and after (dark blue) the onset of star formation in the outer disk. The thin and thick lines indicate the location of 1 and 10\% of the gas mass, respectively. The histograms show the normalized mass-weighted distributions of each quantity. The horizontal line marks the density threshold for star formation. The volume considered is chosen to avoid any contamination from the ISM of the companion galaxy, and to show only the gas of \vintergatan.}
\label{fig:radialpdf}
\end{figure}

\fig{radialpdf} shows the increase of the gas density in the outer galaxy, focusing on the central $10 \kpc$ to avoid contamination by the ISM of the companion galaxy. Before the interaction (light blue lines on \fig{radialpdf}), star forming gas is exclusively found in the innermost region of the galaxy. At the instant pictured in \fig{compress} (medium blue lines in \fig{radialpdf}), tidal compression increases the density in the outer disk ($\gtrsim 7 \kpc$) by several orders of magnitude. The gap in the density profile found at intermediate radii ($\approx 6\mh 7 \kpc$) demonstrates that the increase of density is not a growth of the disk, nor the ejection of tidal debris, but rather a temporary and localized effect in between the galaxies, as illustrated in \fig{compress}.

In the course of the interaction, the separation distance between the two galaxies varies constantly, such that tidal compression eventually spans a wide range of galacto-centric radii (over all azimuths in the outer disk). When the companion reaches its apocenter ($\approx 8.8 \Gyr$ ago, i.e. just before falling back toward final coalescence which occurs $8.7 \Gyr$ ago), high densities are found at all radii in the outer disk (dark blue lines in \fig{radialpdf}).

\begin{figure}
\centering
\includegraphics{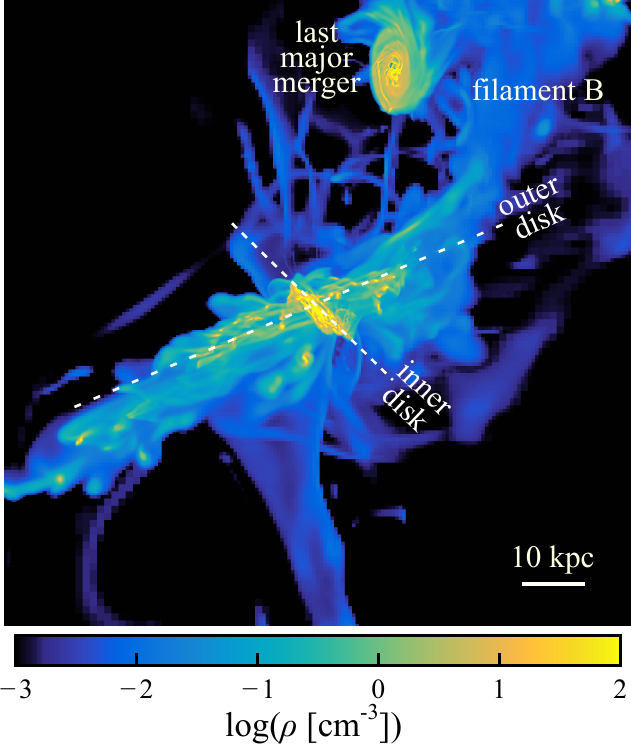}
\caption{Gas density map $\approx 100 \Myr$ before the final coalescence of the two galaxies. The line of sight, different from \fig[s]{tilt} and \ref{fig:compress} is chosen to show the misalignment of the inner and outer disks, and the star forming clumps in the outer structure, of which formation has been tidally triggered earlier during the galactic encounter.}
\label{fig:tilt_00045}
\end{figure}

The probability distribution function of gas density (right panel of \fig{radialpdf}) shows that the global maximum density increases during the interaction, but only by a factor of a few. The main effect is rather a compression of the low density gas ($\lesssim 10^{-1}\cc$, from the diffuse outer disk) toward intermediate and high densities ($\sim 10^{0\mh 4}\cc$). This compression is strong enough to lead to the fragmentation of gas into clumps (\fig{tilt_00045}), to their collapse, and finally to the onset of star formation in the outer disk.

\begin{figure}
\centering
\includegraphics{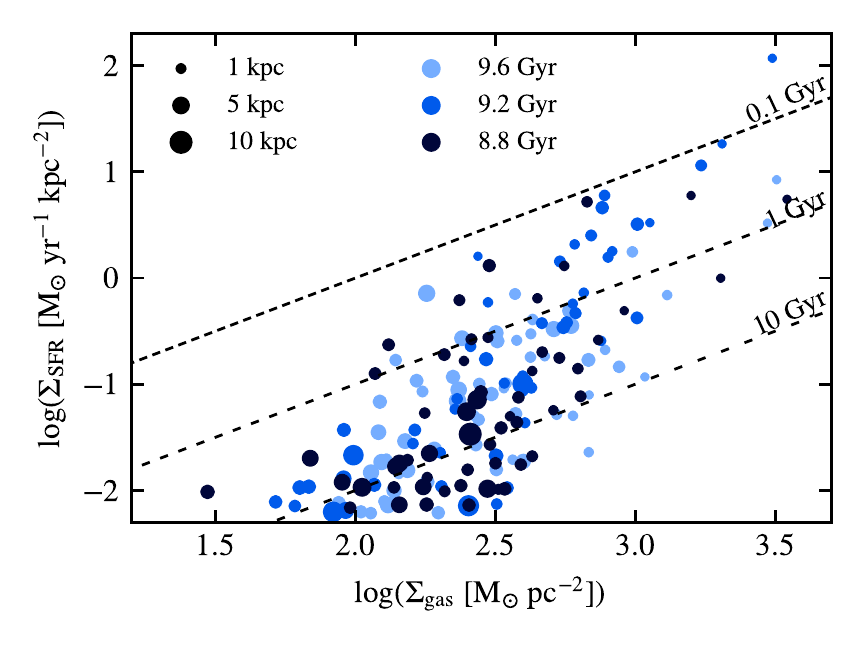}
\caption{Surface density of star formation rate, as a function of the surface density of gas, in bins of $0.5 \kpc\times 0.5\kpc$, at the three epochs considered in \fig{radialpdf}, i.e. before (light blue), during (medium blue) and after (dark blue) the onset of star formation in the outer disk. The size of the symbols indicates the galacto-centric distance. As before, only regions within the innermost $10 \kpc$ are considered here to avoid contamination by the companion galaxy. The dashed lines indicate constant depletion times ($=\Sigma_{\rm gas}/\Sigma_{\rm SFR}$) of 0.1, 1 and $10 \Gyr$.}
\label{fig:ks}
\end{figure}

This only mild modification of the densest end of the density distribution is in contrast with the excess of dense gas induced by tidal compression in starburst galaxies (see e.g. \citealt{Renaud2014b, Renaud2019b}). Therefore, although the global SFR increases during the interaction (by a factor $\approx 2$, see \citetalias{Renaud2020}), it does not globally reach the level of starbursts before the enhancement of the star formation activity in the inner disk at coalescence. This is further illustrated in the Schmidt-Kennicutt diagram in \fig{ks}. The first passage decreases the depletion time ($=\Sigma_{\rm gas} / \Sigma_{\rm SFR}$) in regions which were already forming stars, notably in the galactic center (due to tidal compression and torques leading to nuclear inflows, see \citealt{Keel1985, Renaud2019b}, as visible in \fig{ks} with the small points already at high $\Sigma_{\rm gas}$ moving to higher $\Sigma_{\rm SFR}$). This activity then slows down during the separation phase. 

The compression of gas in the outer disk leads to the onset of star formation in this volume. Yet, the vast majority of these outer regions (large points on \fig{ks}) do not show depletion times as short as in the inner galaxy (small points). At the peak of the star formation activity (medium and dark blue symbols), only regions in the inner disk reach the starburst regime indicated with short depletion times ($\lesssim 1 \Gyr$). This radial dependence of the timescales of star formation translates into different chemical enrichments. With higher $\Sigma_{\rm SFR}$ and shorter depletion times, stars in the inner galaxy release higher-\afe gas into the interstellar medium than those in the outer disk (see e.g. \citealt{Matteucci2012} and \citetalias{Renaud2020}). Therefore, the pre-existing chemical differences noted in \sect{before} and these different enrichments with the onset of star formation imply that the outer disk forms its stellar population with significantly different abundances from that of the inner disk: lower \afe and lower \feh.

Once the interstellar medium is partly fragmented, cooling becomes more efficient and more gas condensates on the clumps (\citetalias{Agertz2020}). The mass of the outer disk increases and star formation continues, even after the initial compressive trigger has stopped.

We note that this star formation process is exclusively in situ, and is not polluted by neither the gaseous nor the stellar contents of the companion galaxy: here, the role of the companion is purely gravitational and remote.

\subsection{Kinematics signatures in the present-day galaxy}
\label{sec:after}

Gravitational torques between the inner and outer disks eventually align the two structures within $\approx 3 \Gyr$ after the onset of star formation in the outer disk (Figure 8 of \citetalias{Agertz2020}). However, the weak gravitational attraction from the outer disk and the (distant) inner galaxy combined with a large intrinsic velocity dispersion of the stars induce large scatters in the kinematic properties of the population formed in this structure. Due to its dissipative nature and the absence of a maintained external excitation, the outer gas disk remains relatively thin while it aligns with the inner disk, and continues to form stars. As a result, the global disk of \vintergatan grows radially (figure 7 of Paper I). The stars formed in the process described here retain an observable signature of their peculiar origin, still detectable at the end of the simulation.

To analyze the signatures of the tilting disk scenario in the present-day galaxy, we distinguish the population of stars formed in the tilted outer disk (selected with $\feh < 0$ and with an in situ formation between 7.7 and $9.2 \Gyr$ ago), the stars of the same age, and those from the low-\afe sequence\footnote{We select these stars as \afe $< 0.39$, independently of their age and metallicity, for simplicity. See however \citet{Hayes2018} advocating for a more precise selection in the real Milky Way.} (most of which having younger ages, as explained below). The high inclination of the outer disk naturally implies that its stars yield a significantly more vertically-extended distribution than the other populations, which both show flatten geometries (in the reference frame of the disk at the end of the simulation).

\begin{figure}
\centering
\includegraphics{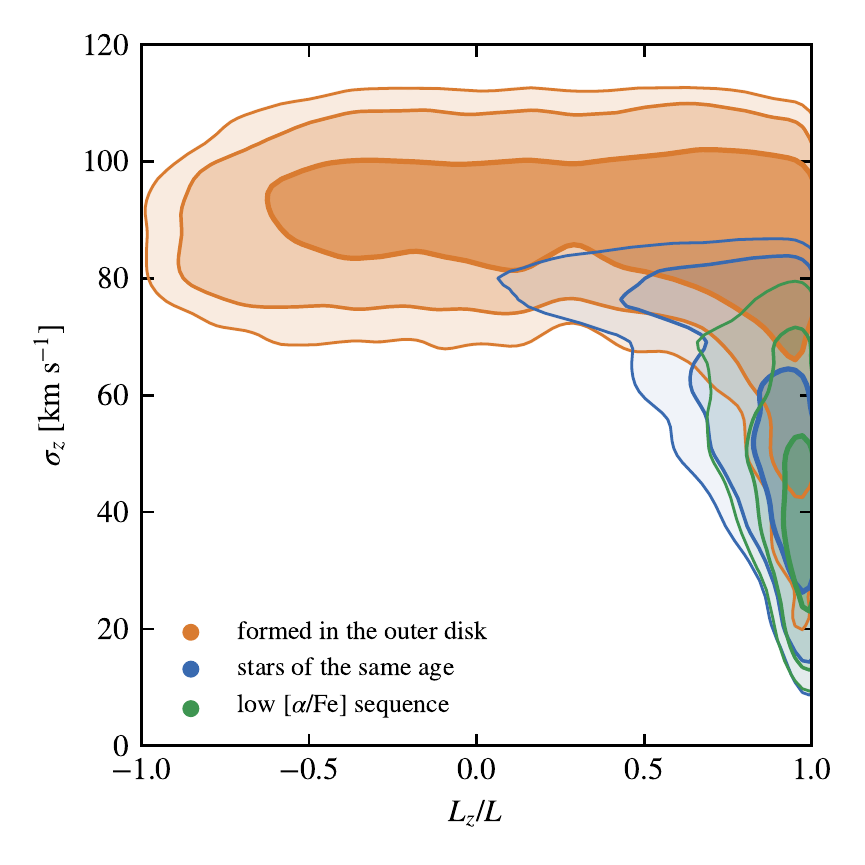}
\caption{Distribution of stars in the plane of vertical velocity dispersion and ratio of the vertical component of the angular momentum to total. All quantities are measured in the final snapshot of our simulation in the cylindrical reference frame of the galaxy. We distinguish the stars formed in the tilting outer disk (orange), the other stars formed in the galaxy at the same epoch (blue), and the population of the low-\afe sequence (green). The contours correspond to 1, 10 and 20 percents of each population. The stars formed in the outer disk (orange) yields halo-like kinematics, reaching high vertical velocity dispersions, and with a large fraction of retrograde orbits. Some stars of the same age (blue) are formed during the starburst episode triggered by the last major merger, thus with tidally-excited kinematics.}
\label{fig:kinematics}
\end{figure}

\fig{kinematics} shows the distribution of the stars in the plane of vertical velocity dispersion $\sigma_z$ and vertical component of the angular momentum $L_z$ normalized by the total momentum $L$ (with all quantities measured at the end of the simulation, in galactic cylindrical coordinates). Not surprisingly, stars from the outer disk yield a significantly higher vertical velocity dispersion, on average, than the other populations. Furthermore, before the inner and outer disks align, the high inclination of the outer disk allows its stars with large velocity dispersions to reach retrograde orbits (with respect to the main galactic axis). Stars on retrograde orbits (30\% of the population from the outer disk, compared to 8\% for the entire galaxy) are also found on a thicker distribution than their counterparts on prograde orbits, indicating they are subject, on average, to a weaker gravitational influence of the disk, which thus allows them to be dynamically hotter.

\begin{figure}
\centering
\includegraphics{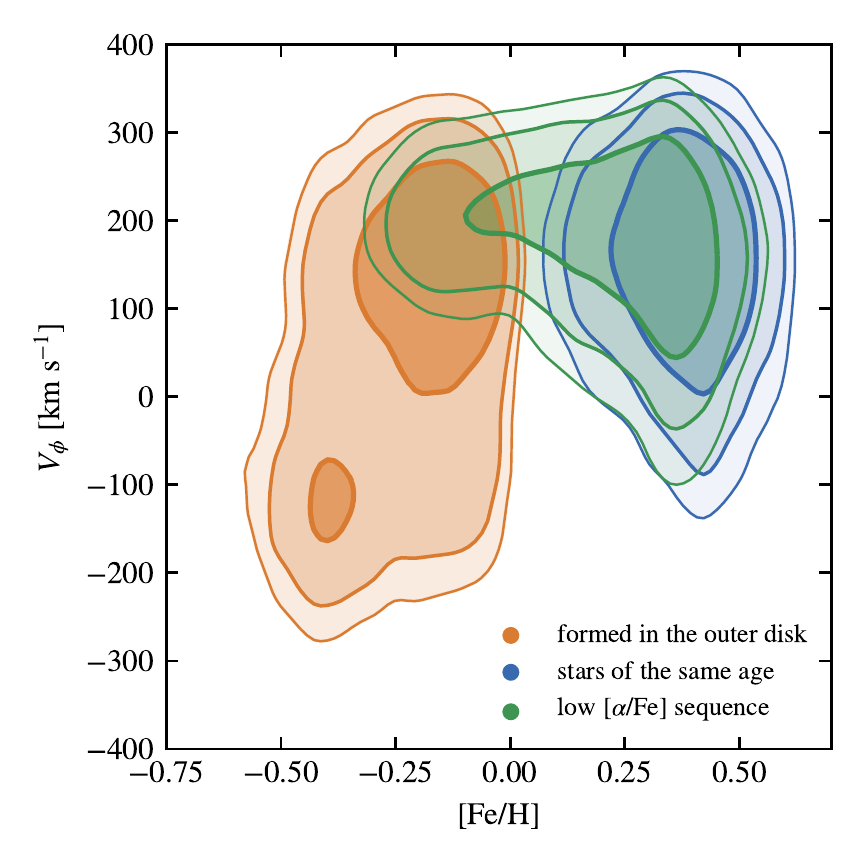}
\caption{Distribution of same populations as in \fig{kinematics}, in the plane of tangential velocity $V_\phi$ and metallicity. The contours correspond to 1, 10 and 20 percents of each population.}
\label{fig:fevtan}
\end{figure}

\fig{fevtan} shows the distribution of our populations in the plane of metallicity and tangential velocity (again in galactic cylindrical coordinates at the end of the simulation, with positive $V_\phi$ corresponding to prograde orbits). As expected, the population from the outer disk stands out in metallicity. The majority of this population is found on prograde orbits and resembles a metal-poor analogue of the stars formed at the same epoch. However, as noted before, a significant fraction of these stars have retrograde motions, which is not found in the equivalent more metal-rich groups. The slow alignment of the two disks progressively makes the intrinsic velocity dispersion of the stars in the outer disk insufficient to reach retrograde orbits (with respect to the main galactic disk). Therefore, while the inclination of the outer disk decreases, an increasing fraction of its stars yield prograde motions. At the same time, the outer disk slowly self-enriches, such that the most metal-poor stars there are the oldest\footnote{This does not apply to the inner disk at this epoch, see \citetalias{Renaud2020}.}, and are the most prone to be found with retrograde motions. This appears as the correlation between $V_\phi$ and \feh visible in \fig{fevtan}.

In short, the initially organized motions of the stars formed in the outer disk is rapidly blurred by their large velocity dispersion, the weak gravitational potential in these volumes, and the slow alignment of the gas disk with the plane of the inner disk. After several Gyr of evolution, the resulting stellar structure constitutes an important fraction of the inner halo of the galaxy ($\lesssim 15 \kpc$) and the now-aligned outer disk, which thus seamlessly connects to the inner thin disk. This process accentuates the inside-out and upside down nature of the formation of the galaxy (\citealt{Bird2013}, see also \citetalias{Agertz2020}).

\section{Discussion}
\label{sec:discussion}

\subsection{The non-accreted origin of low-metallicity stars with halo-like kinematics}

If one would consider spatial, kinematic and metallicity properties \emph{only}, the low-metallicity stars discussed in this paper would seem to point toward an accreted origin, likely a superposition of several accreted satellite galaxies to explain the large range of tangential velocities. However, this is not the case here, since all these stars are formed in situ in the outer disk.

\begin{figure}
\centering
\includegraphics{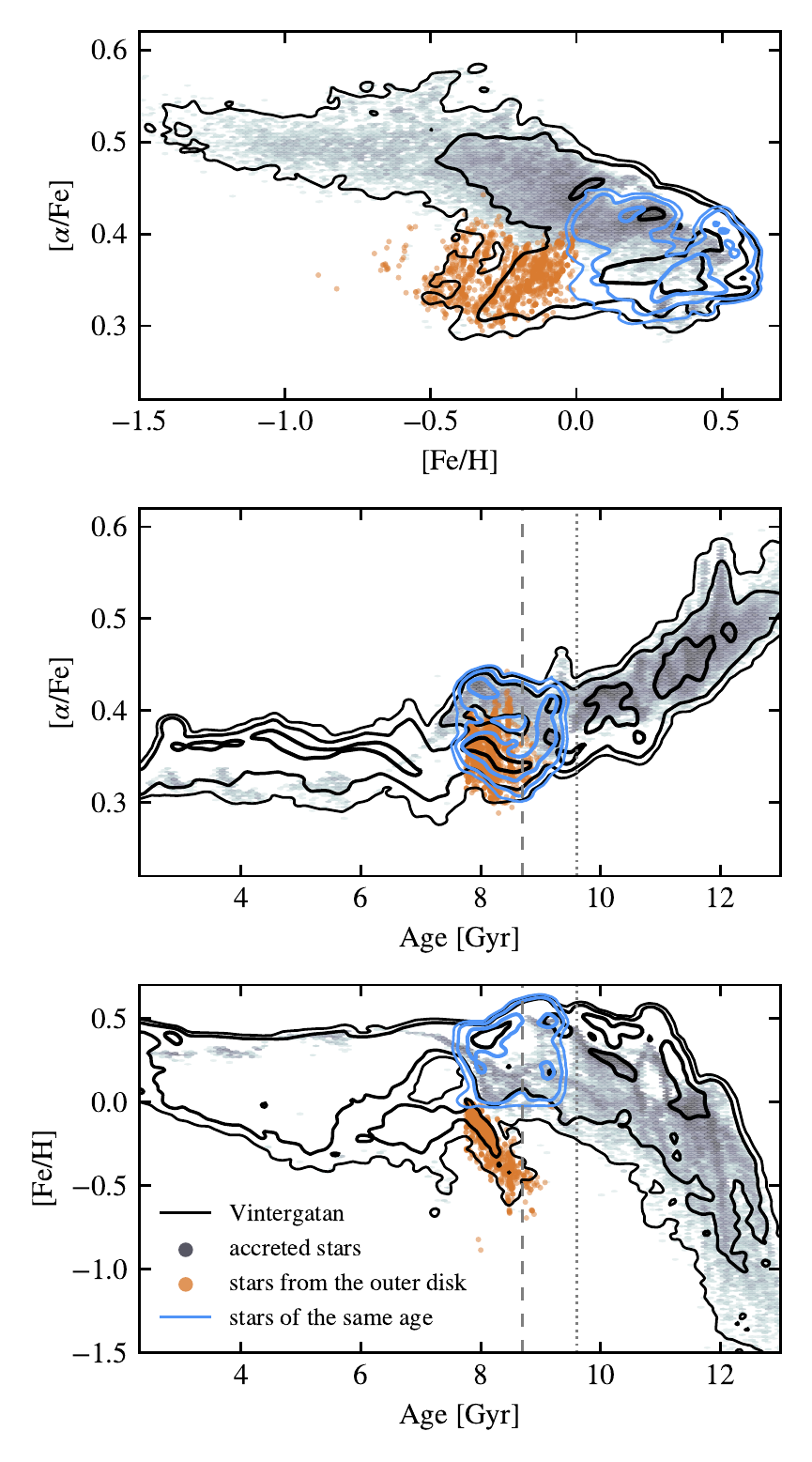}
\caption{Distribution of stars in abundance and age space, showing all the \vintergatan's stars (black contours), the accreted stars (grey-shaded histogram), the population from the outer disk (orange points) and stars formed at the same epoch (blue contours). The contours represent 0.1, 1 and 10\% of their respective populations. The vertical lines indicate the first passage and the final coalescence of the last major merger. The stars from the outer disk lie at the low metallicity-end the low-\afe sequence. As noted above, the first stars born in the outer disk are formed during the interaction phase of the two galaxies.}
\label{fig:age_notlastmm}
\end{figure}

The key resides in the connection of the outer disk population with the thin disk (traced by the bulk of the low-\afe sequence), as illustrated in \fig{age_notlastmm}. Since star formation in the tilting disk is triggered by the last major merger (i.e. not to be followed by any other significant interaction, by definition), it coincides with the end of the starburst phase of galaxy formation, and the onset of the secular, less violent phase (see \citetalias{Renaud2020}). This modification in the star formation regime induces a change in the relative rate of type-II and type-Ia SNe, and thus leads to a decrease in \afe \citep{Wyse1988, Matteucci2012}. Therefore, the stars formed by compression in the tilting disk are among the first to populate the low-\afe sequence, which they do from its low-metallicity end. In the more metal-rich regions of the inner galaxy, shocks and nuclear inflows induced by the late stages of the interaction of the last major merger sustain a regime of short depletion times until final coalescence (as hinted by a large number of the small blue dots at high $\Sigma_{\rm gas}$ and high $\Sigma_{\rm SFR}$ in \fig{ks}, see also \citealt{Renaud2019b}). This slightly delays the decrease of \afe at the high-metallicity end for the duration of the galactic interaction, i.e. a few $100 \Myr$ (middle panel of \fig{age_notlastmm}). The difference in metallicity between the inner and outer disks leads to a clear bimodality in \feh for the stars formed at this epoch. Such a bimodality could be detected in the real Milky Way, given small enough uncertainties on the ages (see \citetalias{Renaud2020}, Figure 12). Improving the observational uncertainties on stellar ages, for instance thanks to larger asteroseismology surveys, could help distinguish our scenario from alternatives. It is even possible that the next generations of instruments will provide sufficient precision over galactic-wide populations to constrain not only the family of formation scenarios, but also their fine details, e.g. the time-line of events, their timescale, intensity etc.

As they align, the inner and outer disk pollute each others (via gas flows and galactic fountains, and indirectly via radial migration, see \citetalias{Agertz2020}). Therefore, the low-\afe sequence is assembled from two channels: at low metallicity by the onset of star formation in the tilting disk triggered by the passage of the last major merger, and at high metallicity by the slowdown of the star formation activity due to the end of the merger-induced starbursts after the last major merger. These two channels are thus two facets of the same event, which implies that their almost-simultaneity is not coincidental.

This assembly from two channels necessarily connects the population of the outer disk to the bulk of the thin disk in chemical and age space (\fig{age_notlastmm}). Such a seamlessly connection is very unlikely if considering an accreted origin. The tilting disk scenario provides the possibility that stars at the metal-poor end of the low-\afe sequence observed to have halo-like kinematics \citep[e.g.][]{Nissen2010, diMatteo2019} could be neither accreted material (as observed, but at lower metallicities than those discussed here, and thus with a more ambiguous connection to the inner disk, \citealt{Belokurov2018, Helmi2018, Haywood2018, Hayes2018, Mackereth2018, Myeong2019}) nor the result of disk heating \citep{Purcell2010, McCarthy2012, Minchev2014}. A dynamical heating of the thin disk would explain the low-\afe content and most of the spatial and kinematic properties of the population. However, it is yet unclear why this population would be significantly more metal-poor than the other stars of the same age \citep{Feuillet2019}, and why it would comprise a significant fraction of stars on retrograde orbits (but see \citealt{Qu2011, JeanBaptiste2017} for the possibility of retrograde orbits when disk heating is caused by \emph{minor} mergers). All these aspects are reconciled in the scenario presented here, where the inner halo comprises stars formed in situ in the tilting outer disk, in addition to the well-known deposition by satellite galaxies \citep{searle1978, Read2008, Renaud2017}. 

This scenario is in remarkable agreement with the recent findings of \citet{Ciuca2020} who used machine learning on APOGEE data to improve estimates on stellar ages. They reported the simultaneous formation of the low- and high-metallicity ends of the low-\afe sequence, approximately at the same epoch as the transition from the high- to low-\afe branches. Their identification of the inner and outer disk components in the age and abundance spaces (their Figure 9) validates our tilting disk scenario (derived independently from their results). Yet, one qualitative difference is their identification of the outer disk stars as making the upper part of the \afe distribution of stars of the same age, while we find them to cover an large fraction of the distribution, preferentially at the lower end. This difference could be due to a more vigorous star formation activity in the outer disk in the real Galaxy than in our simulation, that would have enhanced the \afe content. It is also possibly linked to uncertainties in the chemical enrichment model used in simulations (especially the rate of type-Ia SNe), which could alter the shape and position of features in the abundance plane. Other simulations varying the parameters of the last major merger are necessary to infer the range of possible variations in our scenario.


\subsection{Likelihood of this scenario}

Despite the excellent agreement of our results with a number of recent observations (e.g. \citealt{Feuillet2019}, \citealt{Ciuca2020}, \citealt{Miglio2020}),  the scenario presented here could appear as a very precise (and thus rare) chain of events. However, we stress that almost all the steps are natural consequences of each others. For instance, the formation of the outer gas disk with a different chemical composition from the inner galaxy is a consequence of having two distinct filaments fueling the galaxy. The trigger of star formation in this volume, but not at the level of starbursts, comes from the brushing passage of the companion galaxy (which has previously enriched one, and only one, of the filaments). Finally, the alignment of the outer and inner disks is the simple consequence of their mutual gravitational effects. Therefore, the validity of this scenario only requires two conditions to be satisfied.

(i) A massive galaxy must be fueled by two distinct gaseous filaments, one containing another massive galaxy (that will become the last major merger) and one without any massive galaxy, to ensure different chemical pollutions from outflows. Statistical analysis of the connectivity of galaxies from the SDSS catalogs shows that galaxies of the mass of the Milky Way progenitor are very likely found at the intersections of at least two filaments of the cosmic web (\citealt{Kraljic2020}, even at $z\sim 1.5$, Kraljic et al. in preparation). Because the massive galaxy involved is the \emph{last} major merger, the secondary filament cannot comprise another galaxy of similar or greater mass. Therefore, this first condition is very likely satisfied. 

(ii) The galactic interaction must be strong enough to tidally compress the outer galaxy, but also weak enough not to destroy the existing disks. It is difficult to predict which combinations of the mass ratio, the velocity of the encounter and the impact parameter would fulfill this criterion. We note that the interaction in our simulation is rather long ($\approx 900 \Myr$), due to a large impact parameter (i.e. a weak dynamical friction and thus a long separation phase between the pericenter passages), but still compatible with typical values in non-group environments \citep[e.g.][]{diMatteo2008}. A more direct encounter would probably lead to an intense, galactic-wide, starburst activity but would also dramatically disturb the morphology of the disk, forming long tidal tails and possibly transforming the galaxy into an elliptical \citep{Bournaud2007c}. The presence in the Milky Way of a disk made of stars older than the last major merger implies that such transformations have not happened. Therefore, the interaction modeled in our simulation (and similar setups leading to our scenario) are compatible, at least qualitatively, with the real Galaxy (see \citetalias{Agertz2020} for a detailed comparison with observed galactic properties). In addition, we note that a similar \feh bimodality arising at a certain epoch, as in our simulation, is also present in other, independent, simulations of Milky Way-like galaxies (see e.g. the models of \citealt{Sanderson2020} discussed in \citealt{Belokurov2020}, their Figure 12, and Teyssier, priv. comm.). The scenario presented here can thus be seen as a rather generic chain of events occurring naturally in Milky Way-like galaxies which experience a last major merger at a similar epoch. 

Furthermore, tilting disks are commonly found in cosmological simulations. For instance, in IllustrisTNG100, all the 25 galaxies exhibiting a counter-rotating structure show a highly-tilted disk (or similarly a polar-ring geometry) at some epoch in the redshift range $z \sim 0.2 \mh 1.0$ (i.e. $2\mh 8 \Gyr$ ago, \citealt{Khoperskov2021}). This further illustrates that tilting disks comparable to the one in \vintergatan are found independently in other simulations, under a variety of environmental conditions, and therefore do not strongly depend on physical, nor numerical details. However, under which conditions such structures actually influence the chemical bimodality, as in \vintergatan, still needs to be explored. Quantifying statistically the likelihood of our scenario specifically in Milky Way-like systems, and its role in the chemical composition of stars will require cosmological simulations at high enough resolution to capture thin disks and the compression of diffuse gas, and of sufficient number to explore the role of assembly history and environment. This will become available with the next generations of simulations in the forthcoming exascale era.

\section{Conclusion}
\label{sec:conclusion}

Using the \vintergatan cosmological zoom simulation of the formation of a Milky Way-like galaxy, we identify a mechanism leading to the in situ formation of stars in a disk structure in the outer galaxy, and with a metallicity up to $0.7 \dex$ lower than disk stars formed at the same epoch. This tilting disk scenario requires that the outer galaxy is first fueled with low-metallicity gas that makes an outer disk initially highly inclined. The close passage of the galaxy involved in the last major merger (a few $100 \Myr$ before coalescence) triggers the tidal compression of this gas, which leads to star formation in the outer galaxy. Star formation in this region occurs with long depletion times, therefore maintaining a low \afe. The companion galaxy only acts as a gravitational trigger but does not directly pollute the outer disk, neither in gaseous nor stellar forms. The initially high inclination of the outer disk imparts halo-like kinematics with thin disk-like \afe content to its stars: they have a lower metallicity and are kinematically hotter than the other disk stars of the same age and formed later. However, they have a significantly lower \afe than accreted material. With time, gravitational torques align the outer disk with the inner one, making one large (thin) disk. The tilting disk scenario provides an explanation for the detection of stars with kinematics compatible with an accreted origin, and a thin disk-like chemical composition. The key in matching both, apparently incompatible, properties resides in the evolution of the tilting disk, as a halo structure progressively becoming the outer part of the thin disk.

According to this tilting disk scenario, the galaxy populates its low-\afe sequence in situ, and from two channels: one at high metallicity with a high but decreasing \afe marking the transition from thick to thin disk, and the one presented above at low \afe and low but increasing metallicity. The assembly of the low-\afe sequence from two independent channels is in sharp contrast with other formation scenarios which rely on a sequential assembly (e.g. the two-infall model \citealt{Chiappini1997, Spitoni2019}, the self-enrichment in massive gas clumps \citealt{Clarke2019}, the intrinsic evolution of the star formation activity \citealt{Khoperskov2020}, and the radial migration model \citealt{Schoenrich2009, Minchev2018}). Detecting a bimodality in metallicity for the ages corresponding to the epoch of the last major merger would help distinguishing between the two classes of scenarios, but, observational uncertainties on the stellar ages make this task challenging.

\section*{Acknowledgements}
We thank Sergey Khoperskov for interesting input, and the referee for their report. FR, OA, EA and MR acknowledge support from the Knut and Alice Wallenberg Foundation. OA acknowledges support from the Swedish Research Council (grant 2014-5791). TB was funded by the grant 2018-04857 from the Swedish Research Council. DF was supported by the grant 2016-03412 from the Swedish Research Council. 

\section*{Data availability}
The data underlying this article will be shared on reasonable request to the corresponding author.

\bibliographystyle{mnras}
\bibliography{biblio}

\end{document}